\documentclass[a4paper,runningheads]{llncs}
\usepackage{butterma}
\idline{This is a post-peer-review, pre-copyedit version of an article published in Lecture Notes in Computer Science Vol. 6627. The final authenticated version is available online at: \url{https://doi.org/10.1007/978-3-642-21210-9_22}.}
\usepackage[utf8]{inputenc}
\usepackage{graphicx}
\usepackage{url}
\usepackage[nocompress]{cite}
\newcommand{\code}[1]{{\tt #1}}
\usepackage{hyperref}
\hypersetup{
	bookmarks=true,         
	unicode=false,          
	pdftoolbar=true,        
	pdfmenubar=true,        
	pdffitwindow=false,     
	pdftitle={The Role of Models and Megamodels at Runtime},
	pdfauthor={Thomas Vogel and Andreas Seibel and Holger Giese},
	pdfsubject={},
	pdfcreator={},
	pdfproducer={},
	pdfkeywords={self-adaptive software, megamodels, runtime models},
	pdfnewwindow=true,
}

\begin{document}
\pagestyle{headings}

\title{The Role of Models and Megamodels at Runtime}
\titlerunning{The Role of Models and Megamodels at Runtime}

\author{Thomas Vogel \and Andreas Seibel \and Holger Giese}
\authorrunning{T. Vogel, A. Seibel, and H. Giese}

\institute{Hasso Plattner Institute at the University of Potsdam\\Prof.-Dr.-Helmert-Str. 2-3, 14482 Potsdam, Germany\\
\email{thomas.vogel@hpi.uni-potsdam.de}
}

\maketitle

\thispagestyle{electronic}

\begin{abstract}
In model-driven software development a multitude of interrelated models are used to systematically realize a software system. This results in a complex development process since the models and the relations between the models have to be managed. Similar problems appear when following a model-driven approach for managing software systems at runtime. A multitude of interrelated runtime models are employed simultaneously, and thus they have to be maintained at runtime. While for the development case \emph{megamodels} have emerged to address the problem of managing models and relations, the problem is rather neglected for the case of runtime models by applying ad-hoc solutions.

Therefore, we propose to utilize megamodel concepts for the case of multiple runtime models. Based on the current state of research, we present a categorization of runtime models and conceivable relations between them. The categorization describes the role of interrelated models at runtime and demonstrates that several approaches already employ multiple runtime models and relations. Then, we show how megamodel concepts help in organizing and utilizing runtime models and relations in a model-driven manner while supporting a high level of automation. Finally, the role of interrelated models and megamodels at runtime is discussed for self-adaptive software systems and exemplified by a case study.
\end{abstract}

\section{Introduction}

According to France and Rumpe, there are two broad classes of models in \mbox{\emph{Model-Driven Engineering}~(MDE)}: \emph{development models} and \emph{runtime models}~\cite{France+Rumpe2008}. Development models are employed during the model-driven development of software. Starting from abstract models describing the requirements of a software, these models are systematically transformed and refined to architectural, design, implementation, and deployment models until the source code level is reached.

In contrast, a runtime model provides a view on a running software system that is used for monitoring, analyzing or adapting the system through a causal  connection between the model and the system~\cite{MC.2009.326,France+Rumpe2008}. Most approaches,~like~\cite{GarCHSS04,MBJFS09}, employ \emph{one} causally connected runtime model that reflects a running system. While it is commonly accepted that developing complex software systems using \emph{one} development model is not practicable, we argue that the whole complexity of a running software system cannot be covered by one runtime model defined by one metamodel. This is also recognized by Blair et al. who state ``that in practice, it is likely that multiple [runtime] models will coexist and that different styles of models may be required to capture different system concerns''~\cite[p.25]{MC.2009.326}.

At the \emph{2009 Workshop on Models@run.time} we presented an approach for simultaneously using multiple runtime models at different levels of abstraction for monitoring and analyzing a running software system~\cite{VogelNHGB10}. While abstracting from the running system, each runtime model provides a different view on the system since each model is defined by a different metamodel that focuses on a specific concern, like architectural constraints or performance. At the workshop, our approach raised questions and led to a discussion about simultaneously coping with these models since concerns that potentially interfere with each other are separated in different models. For example, any adaptation being triggered due to the performance of a running system, which is reflected by one runtime model, might violate architectural constraints being reflected in a different model. Thus, there exists relations, like trade-offs or overlaps, between different concerns or models, which have to be considered and managed at runtime.

A similar issue appears during the model-driven development of software.~A multitude of development models and relations between those models have to be managed, especially to maintain traceability information and consistency among the models. An example is the \emph{Model-Driven Architecture}~(MDA) approach that considers, among others, transformations of platform-independent to platform-specific models. Thus, different development models are related with each other, and if changes are made to any model, the related models have to be updated by synchronizing these changes or repeating the transformation. In this context, \emph{megamodels} have emerged as one means to cope with the problem of managing a multitude of development models and relations. The term megamodel originates from ideas on modeling MDA and MDE, which basically consider \emph{a megamodel as a model that contains models and relations between those models or between elements of those models} (cf. \cite{BDDFB07,Bezivin_et_al:2003,Bezivin_et_al:2004,Favre04foundationsof}). 

In contrast, the problem of managing multiple models and relations is neglected for the runtime case and to the best of our knowledge there is no approach that explicitly considers this problem beyond ad-hoc and code-based solutions. In this paper, which is a revision of~\cite{VSG10}, we present a categorization of runtime models derived from the current state of research, and conceivable relations between models of the same or different categories. The presented categories and relations demonstrate the role of models at runtime and that multiple interrelated models are already or likely to be used simultaneously at runtime. Based on that, we propose to apply existing concepts of megamodels for managing runtime models and relations. Such an approach provides a high level of automation for organizing and utilizing multiple runtime models and their relations, which supports the domain of runtime system management, for example, by automated impact analyses across related models. Moreover, we especially discuss the conceptual role of interrelated models and megamodels for self-adaptive systems. 

The rest of the paper is structured as follows. Section~\ref{sec:categorization} presents the categorization of runtime models, conceivable relations between models, and the application of megamodel concepts at runtime. The role of interrelated models and megamodels for self-adaptive systems is discussed in Section~\ref{sec:SAS} and exemplified by a case study in Section~\ref{sec:case-studies}. Finally, the paper concludes with Section~\ref{sec:conclusion}.

\section{Models, Relations and Megamodels at Runtime}\label{sec:categorization}

In this section, we present categories of runtime models and conceivable relations between models of the same or different categories. The categorization is derived from literature, primarily the \emph{Models@run.time} workshops~\cite{MRT} and our own work~\cite{GSV-MRT09,VG10,Vogel-ICAC09,VogelNHGB10}. However, we do not claim that the categories are complete or that each category has to exist in every approach. Nevertheless, they indicate the role of models at runtime and demonstrate that different kinds of interrelated runtime models are already or likely to be employed simultaneously.

\subsection{Categories of Runtime Models}\label{subsec:models}

Both of the already mentioned approaches~\cite{GarCHSS04,MBJFS09} employ one runtime model that is causally connected to a running system. In contrast, our approach~\cite{VG10,Vogel-ICAC09,VogelNHGB10} provides multiple runtime models simultaneously, each of which is causally connected to the system and 
specified by a distinct metamodel. Nevertheless, the other approaches also maintain additional model artifacts at runtime, which are not causally connected to a system, but which are used to manage the system. 

In the case of \emph{Rainbow}~\cite{GarCHSS04}, such artifacts are invariants that are checked on a causally connected architectural model, and adaptation strategies that are applied if the invariants are violated. Morin et al.~\cite{MBJFS09} even have in addition to a causally connected architectural model, a feature model describing the system's variability, a context model describing the system's environment, and a so called reasoning model specifying which feature should be activated or deactivated on the architectural model depending on the context model. 

\begin{figure}[t]
\begin{center}
  \includegraphics[width=.73\textwidth]{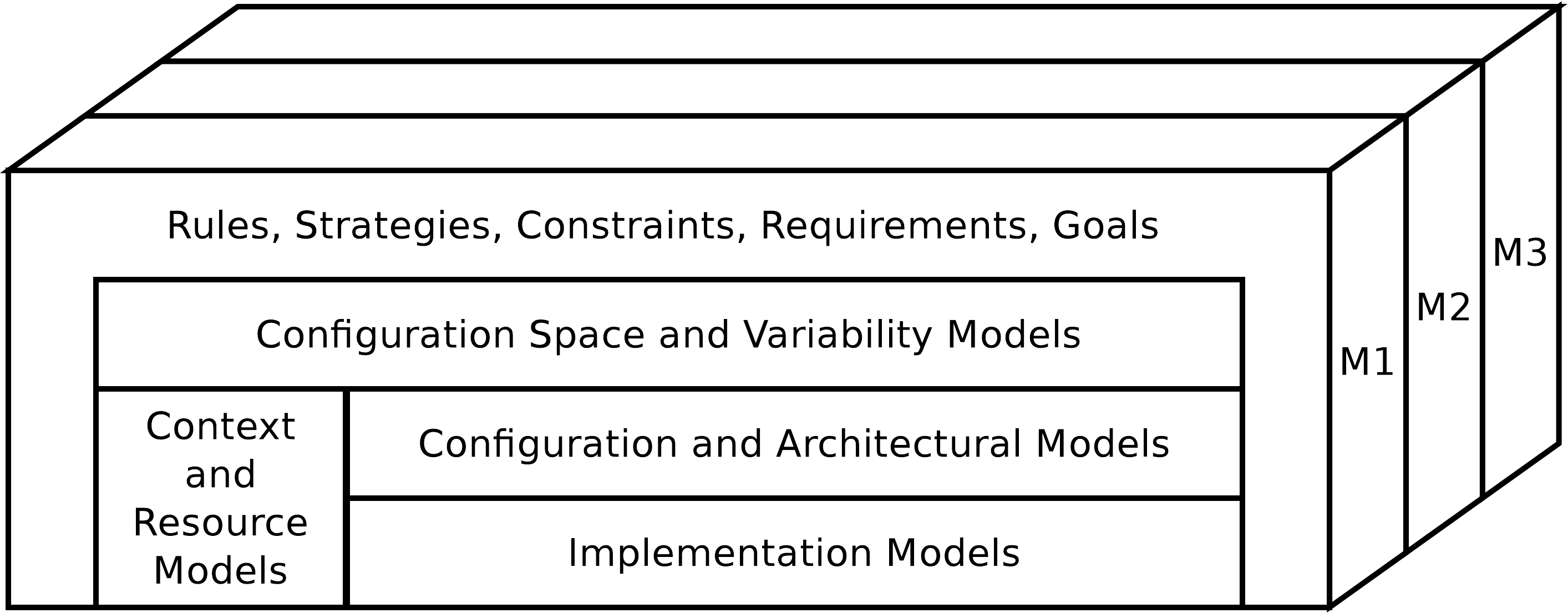}
  \caption{Categories of Runtime Models}
  \label{fig:model-categories}
\end{center}
\vspace{-5mm}
\end{figure}

Thus, even if only one causally connected runtime model is used for managing a running system, several other models that do not need to be causally connected are employed at runtime. For the following categories\footnote{A detailed description of the categories and supporting literature can be found in~\cite{VSG10}.} as depicted in Figure~\ref{fig:model-categories}, we consider any conceivable runtime models regardless whether they are causally connected to a running system or not. The models are categorized in a rather abstract manner according to their purposes and contents. As shown in Figure~\ref{fig:model-categories}, runtime models (\emph{M1}) of all categories are usually instances of metamodels (\emph{M2}) that are defined by meta-metamodels (\emph{M3}), which leverages typical MDE techniques, like model transformation or validation, to the runtime.

\textbf{Implementation Models} are similar to models used in the field of reflection to represent and modify a running system through a causal connection. Such models are based on the solution space of a system as they are coupled to the system's implementation and computation model~\cite{MC.2009.326}. Therefore, these models are platform-specific and at a rather low level of abstraction. As modeling languages, class or object diagrams are often employed to provide structural views, and sequence diagrams or automatons for behavioral views.

\textbf{Configuration and Architectural Models} are usually also causally connected to a running system and they reflect the current configuration of the system. Since software architectures are considered to be at an appropriate abstraction level for analysis or adaptation, these models provide architectural views similar to component diagrams~\cite{GarCHSS04,MBJFS09,VG10,VogelNHGB10}. These diagrams are often enhanced with non-functional properties to directly support analysis or to transform them to specific analysis models, like queueing networks to reason about the performance. At a higher level of abstraction, process or workflow models are also feasible to describe a running system from a business-oriented view. Moreover, model types of the \emph{Implementation Models} category are also conceivable in this category, but at a higher level of abstraction. For example, a sequence diagram would consider the interactions between component instances instead of the interactions between objects.

In general, models of this category and \emph{Implementation Models} are often both causally connected to a system. However, \emph{Configuration and Architectural Models} are at a higher level of abstraction, less complex and often platform-independent, while \emph{Implementation Models} are at a lower level of abstraction, more complex and platform-specific. Thus, \emph{Configuration and Architectural Models} are rather related to problem spaces, and \emph{Implementation Models} to solution spaces. This is similar to the view of Blair et al.~\cite{MC.2009.326} on runtime models and reflection models.

\textbf{Context and Resource Models} describe the operational environment of a running system. This comprises the context, which is ``any information that can be used to characterise the situation of an entity'', while ``an entity is a person, place, or object that is considered relevant to the interaction between a user and an application''~\cite[p.5]{593572} or in general to the operation of the application. To represent a context, semi-structured tags and attributes, key value pairs, object-oriented or logic-based models, or even feature models can be used. Moreover, the operational environment consists of resources a running system requires and actually uses for operation. These are logical resources, like any form of data, or physical resources, like the hardware the system is running on.

\textbf{Configuration Space and Variability Models} specify potential variants of a system, while \emph{Configuration and Architectural Models} reflect the currently running variant of the system. Therefore, models of this category describe a system at the type level to span the system's configuration space and variability. Using these models, adaptation points in a running system and possible adaptation alternatives can be identified. Examples for models in this category are aspect and feature models~\cite{MBJFS09}, or component type diagrams~\cite{GSV-MRT09,VG10}.

\textbf{Rules, Strategies, Constraints, Requirements and Goals} may refer to any model from the other categories and, therefore, their levels of abstraction are similar to the levels of the referred models. Models in this category define, among others, when and how a running system should be adapted by following one of two general approaches. First, rules or strategies usually in some form of event-condition-action rules describe when and under which conditions, a system is adapted by performing reconfiguration actions. The second approach is based on goals a running system should achieve, and guided by utility functions, adaptation aims at optimizing the system with respect to these goals.

Moreover, constraints on models of the other categories regarding functional and non-functional properties are used for runtime validation and verification. Constraints can be expressed, among others, in the \emph{Object Constraint Language}~(OCL)or formally in some form of \emph{Linear Temporal Logic}~(LTL). Though constraints can be seen as requirements that are checked at runtime, recently the idea of \emph{requirements reflection} has emerged, which explicitly considers requirements as adaptive runtime entities~\cite{1810329}. Thus, requirements models, like goal models, become runtime models above the abstraction level of architectures.

The presented model categories show that different aspects have to be considered for managing a system at runtime. These aspects are at least the running system at different levels of abstraction, the system's environment, the system's variability, and the validation, verification and adaptation. Rather than covering all these aspects in a monolithic runtime model being highly complex, multiple and different kinds of models are possible, and even employed simultaneously for that. Which categories and especially which kind of and how many models are employed is specific to each approach. This depends, among others, on the purposes of an approach and on the domain of the system. Nevertheless, separating aspects in different models requires to consider relations among these models.

\subsection{Relations Between Runtime Models or Model Elements}\label{subsec:relations}

In the following, we use the presented model categories to outline exemplars of relations between runtime models or between elements of different runtime models. Note that a relation between elements of two different models also constitutes a more abstract relation between these two models. These exemplars motivate the need for managing relations together with the models at runtime.

As already mentioned, models of the category \emph{Rules, Strategies, Constraints, Requirements and Goals} may refer to models of the other categories. For example, goal modeling approaches refine a top-level goal to subgoals recursively until each subgoal can be satisfied by an agent being a human or a component (cf.~\cite{1810329}). Having a goal model at runtime, it is of interest which component of a running system actually satisfies or fails in satisfying a certain goal. Therefore, goals being reflected in a goal model refer to corresponding components of \emph{Configuration and Architectural Models}, which also relates the goal and architectural model with each other. Moreover, goal satisfaction can be influenced by the current context of a system, such that goals and elements of a context model are related with each other. As an example, consider an e-mail client application that has the subgoal of actually sending a message to a mail server for distribution. This subgoal is fulfilled by a client component that establishes a connection to the server and transmits the message. Thus, this subgoal and component are related with each other. Moreover, satisfying this subgoal is influenced by the availability of a network connection to the server, which is part of the context. This constitutes a relation between the goal model and the context or resource model.

\emph{Configuration and Architectural Models} can also be related to \emph{Configuration Space and Variability Models} by means of effects the selection of a variant as defined by a variability model has on the current system configuration or architecture. For example, activating or deactivating features in a feature model specifying the system's variability requires the adaptation of the currently active architecture by adding or removing corresponding components. Thus, components and their supported features are related with each other. Regarding the same dimension of abstraction, \emph{Implementation Models} can be seen as refinements of \emph{Configuration and Architectural Models} as they describe how a configuration or architecture is actually realized using concrete technologies. Thus, refinement relations are conceivable between models of these two categories.

Another relation can reflect the deployment or resource utilization of a system by means of relating \emph{Architectural Model} elements and \emph{Resource Model} elements, or in other words, which components of a running systems are deployed on which nodes and are consuming which resources. \emph{Context and Resource Models} can also refer to \emph{Configuration Space and Variability Models} since the configuration space and variability of a system can be influenced by the current context or resource conditions. For example, a certain variant is disabled due to limited resources.

Besides relations between models of different categories, there may also exist relations between models of the same category. In~\cite{VogelNHGB10}, several \emph{Architectural Models} are employed reflecting the same system, but providing different views on it. However, these views are overlapping, which can be considered as a relation. Furthermore, each model focuses on a certain concern, like performance or architectural constraints, and any adaptation optimizing one concern might interfere with another concern. As an example, due to a decline in the system performance, an additional component of a certain type should be deployed to balance the load, which however violates the architectural constraint restricting the number of deployed components of the specific type. Thus, overlaps, trade-offs or conflicts between concerns respectively between the models are conceivable.

The presented exemplars show that runtime models are usually not isolated and independent from each other, but they rather compose a network of models. Therefore, besides the runtime models also the relations between those models have to be managed at runtime. The concrete relations emerging in an approach depend, among others, on the purposes of the approach, the domain of the system and on the models that are employed.

\subsection{Megamodels at Runtime}\label{subsec:megamodels-at-runtime}

As it turned out, different kinds of models and relations between them emerge when managing a system at runtime. In such scenarios, it is important that these relations are modeled and maintained at runtime because this makes the relations \emph{explicit} and, therefore, amenable for analysis or other runtime activities. For example, an impact analysis is leveraged when knowing which models are related with each other. Then, the impact of any model change to related models can be analyzed by following transitively the relations and propagating the change. Moreover, relations can be classified, for example in critical and non-critical ones, and for certain costly analyses only the critical relations may be considered.

Nevertheless, relations to other models are usually not covered by all models because they were not foreseen when designing the corresponding modeling languages. Thus, a language for explicitly specifying all kinds of relations between various models and between elements of different models is required. Rather than applying ad-hoc and code-based solutions to relate models with each other, megamodels provide a language that supports the modeling of arbitrary models and relations between those models or between elements of those models. Therefore, the management of models and relations itself is done in a model-driven manner enabling the use of existing MDE techniques for it.

In general, megamodels for the model-driven software development serve organizational and utilization purposes that should also be leveraged at runtime. Organizational purposes are primarily about managing the complexity of a multitude of models. Therefore, like some kind of a model registry~\cite{Bezivin_et_al:2004}, megamodels help in organizing a huge set of different models together with their relations by storing and categorizing them. Likewise, megamodels can serve as a means to explicitly organize and maintain runtime models and their relations in the domain of runtime system management since several models and relations can be employed simultaneously at runtime (cf. Sections~\ref{subsec:models} and~\ref{subsec:relations}).

Utilization purposes of megamodels are primarily about navigation and automation by actually using the relations that are made explicit due to the organizational purposes. Utilizing relations, megamodels can be the basis for navigating through various models. Thus, starting from a model, all related models can be reached in a model-driven manner instead of using mechanisms at a lower level of abstraction like programming interfaces. Having the conceivable runtime models and relations in mind (cf. Sections~\ref{subsec:models} and~\ref{subsec:relations}), navigating between runtime models is essential for a comprehensive system management approach. Thus, explicit relations can be utilized by typical operations to \emph{read} or \emph{write} models, or to \emph{apply} a model on another model. Navigating between models can be considered as reading models, while writing can be a model update by propagating model changes along relations. Finally, models, like transformation or generation rules, can be applied on models resulting in models. This leads to the aspect of automation aiming to increase efficiency. Relations between models are treated as executable units that take models as input and produce models as output. Thus, a megamodel can be considered as an executable process for runtime activities, like automatically analyzing the impact of changes. Therefore, relations can be used to synchronize model changes to related models and these synchronized models are then analyzed to investigate the impact of the initial changes. 

Finally, automation considers the efficient maintenance of models and relations by means of their validity and consistency, because models and relations are often both dynamic and they change over time.

\section{Self-Adaptive Software Systems}\label{sec:SAS}

While in the previous section the model categories are derived from literature and broken down according to the purposes and primarily the contents of the models, now we will approach a different categorization by taking a conceptual view on self-adaptive software systems. Based on the typical feedback loop of a self-adaptive system, we investigate the role of runtime models and especially how they are used throughout the loop. This results in a different model categorization that focuses on the usage of models and that will be compared to the previously presented categorization in order to foster the comprehension of conceivable models and their roles at runtime for self-adaptive systems.

\subsection{Runtime Models for Self-Adaptive Software Systems}\label{subsec:SAS+Models}

\begin{figure}[t]
\begin{center}
  \includegraphics[width=.78\textwidth]{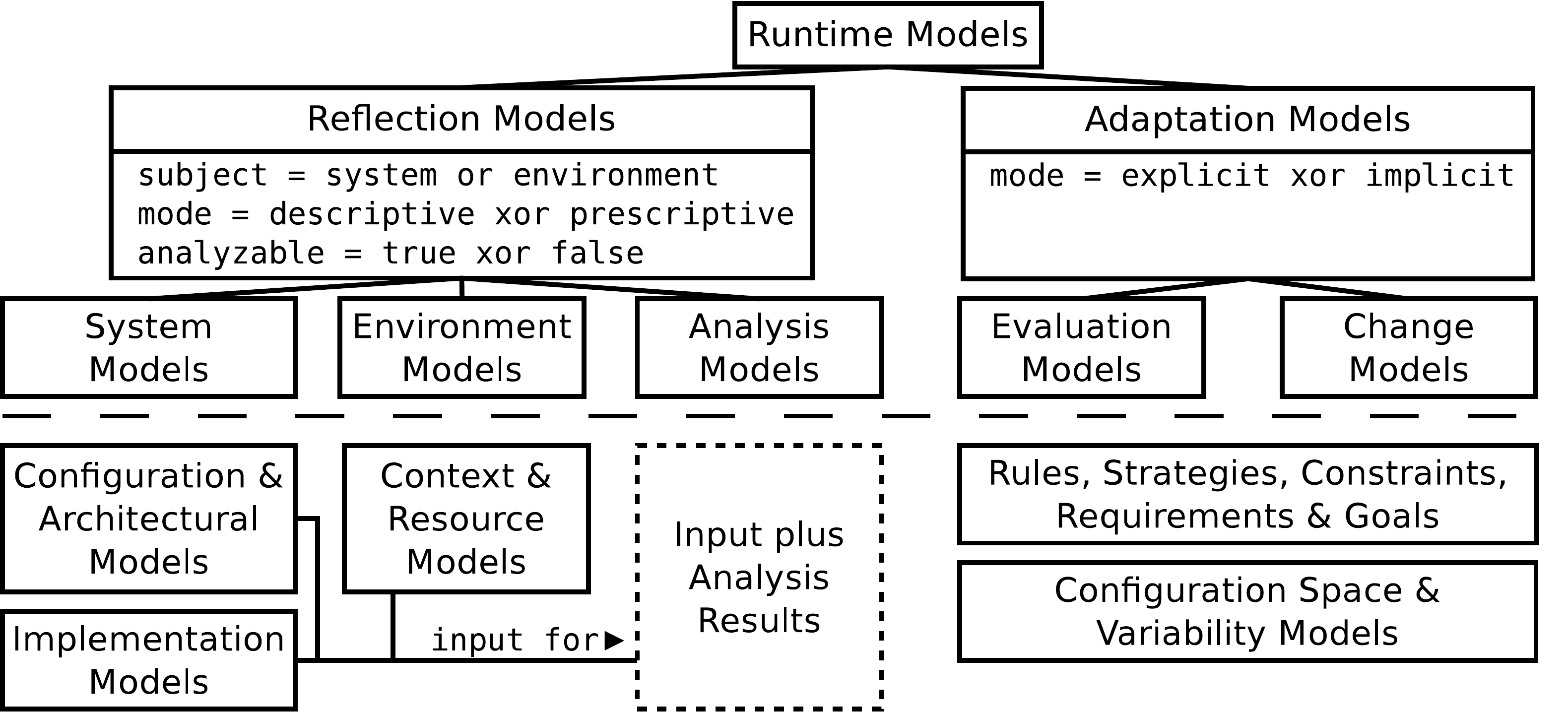}
  \caption{Categories of Runtime Models for Self-Adaptive Software Systems}
  \label{fig:model+loop+categories}
\end{center}
\end{figure}

Before investigating the usage of models in self-adaptive software systems, we describe the categories of these models as depicted on top of Figure~\ref{fig:model+loop+categories}. \emph{Runtime Models} are divided into two top categories, \emph{Reflection} and \emph{Adaptation Models}, based on the way they are used at runtime.

\textbf{Reflection Models} reflect the \code{system} or the system's \code{environment} either in a \code{descriptive} or {\tt prescriptive} manner as indicated by the attributes \code{subject} and \code{mode}, respectively. Descriptive models describe the as-is situation of the running system or environment, while prescriptive models prescribe the to-be situation, primarily the designated target state of the system. Though it is not possible to prescribe the environment, to-be environment models are conceivable for reflecting predictions of the future environment. Reflection models can be \code{analyzable} to support reasoning about the system or environment. Thus, using basic and incomplete attribute assignments, \emph{System Models} (\code{subject = system}), \emph{Environment Models} (\code{subject = environment}), and \emph{Analysis Models} (\code{analyzable = true}) are considered as typical reflection models, while other models are conceivable regarding the possible combinations of the reflection models' attributes. In general, reflection models are primarily \emph{read} and \emph{written} to describe, prescribe or analyze the system and environment.

\textbf{Adaptation Models} on the other hand are primarily \emph{applied} on reflection models as they define how reflection models are evaluated or changed. Thus, \emph{Evaluation Models} specify the reasoning and analyses that are performed on descriptive or prescriptive reflection models, while \emph{Change Models} specify how prescriptive reflection models can be obtained. This can be done either in an \code{explicit} or \code{implicit} \code{mode}. Explicit models enumerate patterns that can be directly compared to reflection models for evaluation or that precisely define fragments of possible prescriptive reflection models. In contrast, implicit models, like rules, define operations that are applied on reflection models, which returns either evaluation results or changed and potentially new reflection models.   

This model categorization can be mapped to the categorization previously presented in Section~\ref{sec:categorization}, which is outlined in Figure~\ref{fig:model+loop+categories}. \emph{System Models} directly correspond to \emph{Configuration \& Architectural Models} and \emph{Implementation Models}, while \emph{Environment Models} are equivalent to \emph{Context \& Resource Models}. In contrast, \emph{Analysis Models} are only implicitly represented in the previous categorization by mentioning that analysis can be performed on \emph{Configuration \& Architectural Models} or on models derived from them. However, this view neglected the important role of the environment for the analysis. Thus, besides models reflecting the system, also environment models have to be considered when creating analysis models. Thus, \emph{Configuration \& Architectural}, \emph{Implementation}, and \emph{Context \& Resource Models} serve as the input for analysis models that also contain the analysis results. Technically, these input models or parts of them can be copied or transformed into the analysis models, or the analysis models can reference the relevant parts of the original input models. A main difference between both categorizations is that the previous categorization does not distinguish between descriptive and prescriptive reflection models.

Finally, \emph{Adaptation Models} can be mapped to \emph{Rules, Strategies, Constraints, Requirements \& Goals} and to \emph{Configuration Space \& Variability Models}. The previous categorization does not clearly distinguish whether the corresponding models are exclusively used for reasoning (\emph{Evaluation Models}) or for specifying and executing changes (\emph{Change Models}). From a conceptual view, applying \emph{Evaluation Models} does not modify \emph{Reflection Models} as they are only read for reasoning purposes, while the application of \emph{Change Models} modify or create new \emph{Reflection Models}, primarily prescriptive \emph{System Models}. However, from a technical view, \emph{Evaluation} and \emph{Change Models} can be quite similar as both can be specified, for example, in some form of rules. 

Moreover, \emph{Configuration Space \& Variability Models} can be especially mapped to \emph{explicit Adaptation Models} as they, for example, explicitly describe potential variants for prescriptive system models. On the other hand, \emph{Rules} or \emph{Strategies} can be mapped to \emph{implicit Adaptation Models} as the prescriptive \emph{System Models} are obtained by sequentially applying the rules or strategies on a descriptive \emph{System Model}. Thus, \emph{explicit Adaptation Models} in the form of \emph{Configuration Space \& Variability Models} are not necessarily required as the adaptation might also be specified by \emph{implicit Adaptation Models} in the form of rules or strategies.

\subsection{Model Operations and Relations for Self-Adaptive Systems}\label{subsec:SAS+Relations}

Using the \emph{Reflection} and \emph{Adaptation Models} and their specializations as described in Section~\ref{subsec:SAS+Models}, and having both categorizations in mind, we outline how a self-adaptive system uses runtime models throughout the feedback loop.

\begin{figure}[t]
\begin{center}
  \includegraphics[width=.84\textwidth]{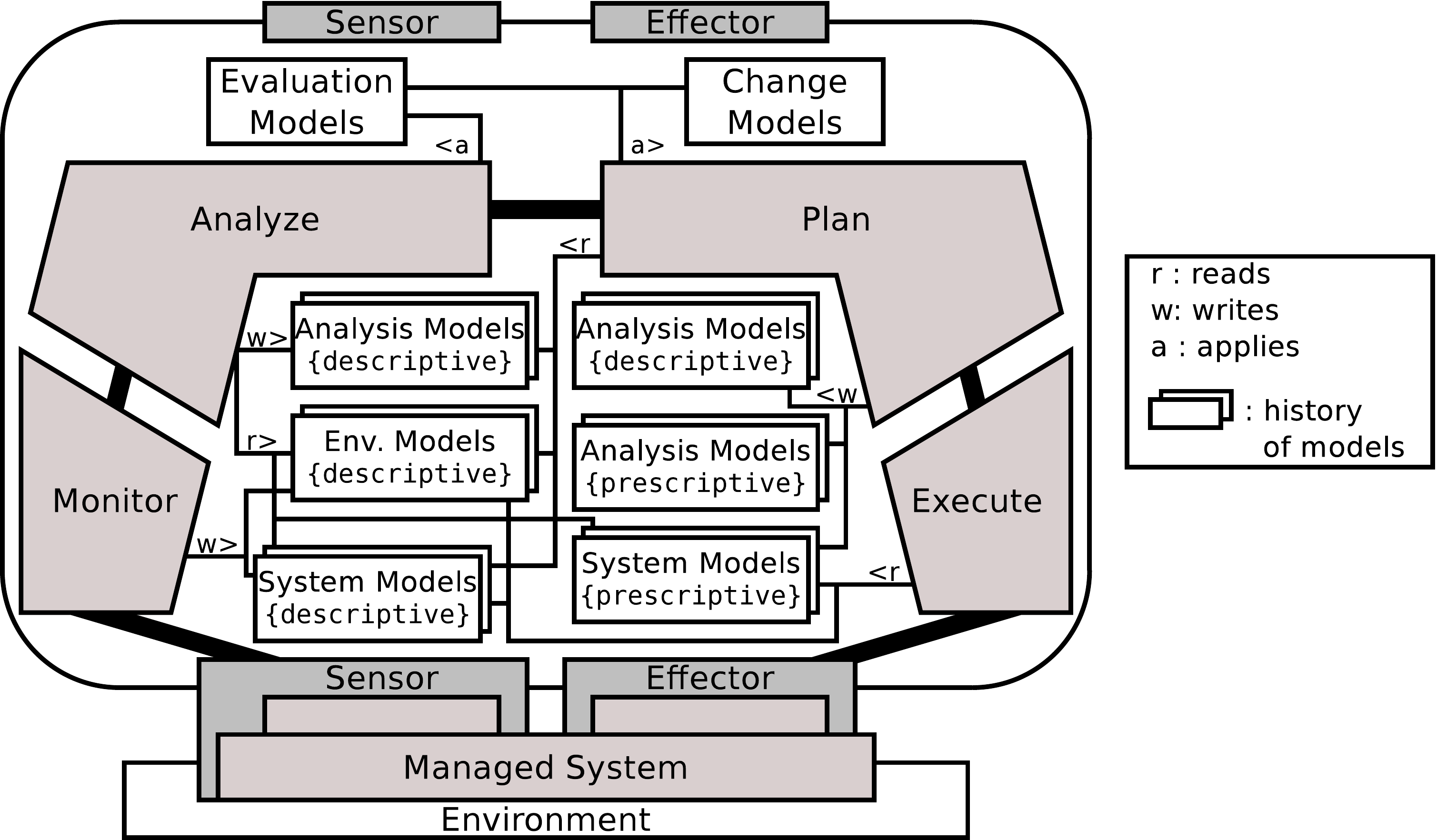}
  \caption{Runtime Models and their Usage along the Feedback Loop}
  \label{fig:model+loop}
\end{center}
\vspace{-5mm}
\end{figure}

The concept of feedback loops is an inherent part of each self-adaptive system since the loop controls the self-adaptation~\cite{Brun+2009}. A generic loop is proposed in~\cite{KephartChess2003} and whose building blocks can be identified in Figure~\ref{fig:model+loop}. The \emph{Managed System} operates in an \emph{Environment} and contains the business logic offered to users or other systems. It provides \emph{Sensor} and \emph{Effector} interfaces to enable its management by autonomic managers implementing the feedback loop. Using sensors, the manager \emph{monitors} and \emph{analyzes} the system and environment to decide whether the system fulfills the given goals or not. If not, adaptation is required and thus, changes are \emph{planned} and \emph{executed} to the system through effectors. A manager itself also provides sensors and effectors to support the hierarchical composition of managers. Additionally, the original loop~\cite{KephartChess2003} considers a generic notion of \emph{Knowledge} that is used and shared by the loop activities. In contrast to the activities, the knowledge remains rather abstract as it is not clearly substantiated. Therefore, we elaborated the role of models for the knowledge by investigating from a conceptual view what types of models are shared and how they are used by the activities. This has lead to the extended loop as shown in Figure~\ref{fig:model+loop}.

Since the loop activities use models to perform their tasks and to communicate with each other, each loop activity can be conceptually considered as a complex and high level model operation taking models as input and producing models as output. Thus, the activities can be seen as relations between the input and output models (cf. Section~\ref{subsec:megamodels-at-runtime}). However, from a technical view the models need not to be completely copied from one activity to another as the same model instances can also be shared among activities or only single model changes, the deltas, can be exchanged between the activities. In the following we discuss \emph{one} reasonable and conceptual scenario for the loop behavior, while considering the loop activities as complex model operations that consist of the basic operations of reading, writing, and applying models. The semantics of these basic operations are substantiated through the application domain of self-adaptive systems. Moreover, the basic operations are the foundation for advanced MDE techniques, like model transformation, synchronization, or merge, being relevant for engineering self-adaptive systems employing runtime models. 

The monitor \emph{writes} \emph{descriptive} \emph{System} and \emph{Environment Models} to continuously provide up-to-date views on the running system and environment, respectively. In general, writing a model includes the \emph{reading} of the model, such that the models do not have to be created every time from scratch but they can also be maintained and updated at runtime. Likewise, a history of models can be maintained to keep track on their evolution over time, like the past states of the system or environment. Moreover, the monitor may filter, merge, abstract, etc. the monitored data to provide several system and environment models simultaneously, that differ, for example, in their abstraction levels.

The descriptive system and environment models are \emph{read} by the analyze step to transform, synchronize or generally \emph{write} them to \emph{descriptive Analysis Models} for reasoning, like a queueing network model for performance analysis. Moreover, \emph{prescriptive System Models} are \emph{read} since they serve as reference models for descriptive system models to analyze whether the current system converges to the designated target state. The analysis itself is defined by \emph{Evaluation Models} that describe implicitly or explicitly the goals of the system in an operationalized form (cf. Section~\ref{subsec:SAS+Models}). Thus, the fulfillment of goals can be analyzed by \emph{applying} evaluation models on system, environment, or analysis models. Based on the analysis results, usually annotated to analysis models, a decision about the need of adaptation is made. If adaptation is required, the planner comes into play.

The planner \emph{reads} the descriptive analysis, system and environment models to devise a plan on how the system should be adapted such that the system fulfills its goals. This planning process is guided by \emph{Change Models} that are \emph{applied} on the descriptive system and environment models to obtain and \emph{write} \emph{prescriptive System Models} reflecting suitable target configurations. Likewise to evaluation models, change models specify implicitly or explicitly how prescriptive system models can be obtained (cf. Section~\ref{subsec:SAS+Models}). Since the planner has to select one among many possible target configurations, analysis is performed to determine the best or at least the most appropriate target configuration with respect to the current system and environment state. Therefore, the planner \emph{reads} and \emph{writes} \emph{descriptive} and \emph{prescriptive} \emph{Analysis Models} by \emph{applying} \emph{Evaluation Models} to reason about the current and the possible future situations. The planning result is a predictive system model describing the final target system configuration.

Finally, this \emph{predictive} and the current \emph{descriptive} \emph{System Models} are \emph{read} by the execute step, and compared with each other to derive the set of reconfiguration actions. These actions move the managed system from the current to the target configuration. Therefore, they are executed on the system through effectors, while considering the latest \emph{descriptive System} and \emph{Environment Models} to find a point in time when the running system can be safely reconfigured.

As already mentioned, an autonomic manager providing sensors and effectors can be managed by another manager, which leads to a hierarchical composition of managers. A higher level manager comes into play when the lower level manager cannot cope with the adaptation of the system, like the planner is not able to find any target configuration fulfilling the goals. Therefore, the higher level manager can perform more sophisticated planning, even at the level of goals, and provide new \emph{Evaluation} and \emph{Change Models} specifying new adaptation mechanisms to the lower level manager. Thus, the higher level manager senses all required models from the lower level one, but it only effects the evaluation and change models and thus, the adaptation mechanisms of the lower level manager. Other triggers for adapting the evaluation and change models of a manager are the emergence of new application or adaptation goals for this manager.

\subsection{Megamodels at Runtime for Self-Adaptive Systems}\label{subsec:SAS+Megamodels}

From the previous sections it can be concluded that different kinds of models are used in different ways throughout the feedback loop of self-adaptive systems. The models are not only used by the loop activities, but they are also shared between the different activities and even between different loops. The relations between models that are described in Section~\ref{subsec:relations} also hold for the case of self-adaptive systems. Moreover, each loop activity can be considered as a complex model operation taking models as input and producing models as output, which similar to the view of megamodels on relations as executable units (cf. Section~\ref{subsec:megamodels-at-runtime}). Thus, the whole feedback loop can be interpreted as an executable process that can be modeled and enacted with a megamodel. By modeling, the comprehension of the feedback loop will be leveraged, and by enacting, the level of automation for executing a loop will be increased through model-driven techniques.

\vspace{-2mm}
\section{Case Study: Self-Adaptive Software Systems}\label{sec:case-studies}

\begin{figure}[t]
\begin{center}
  \includegraphics[width=.94\textwidth]{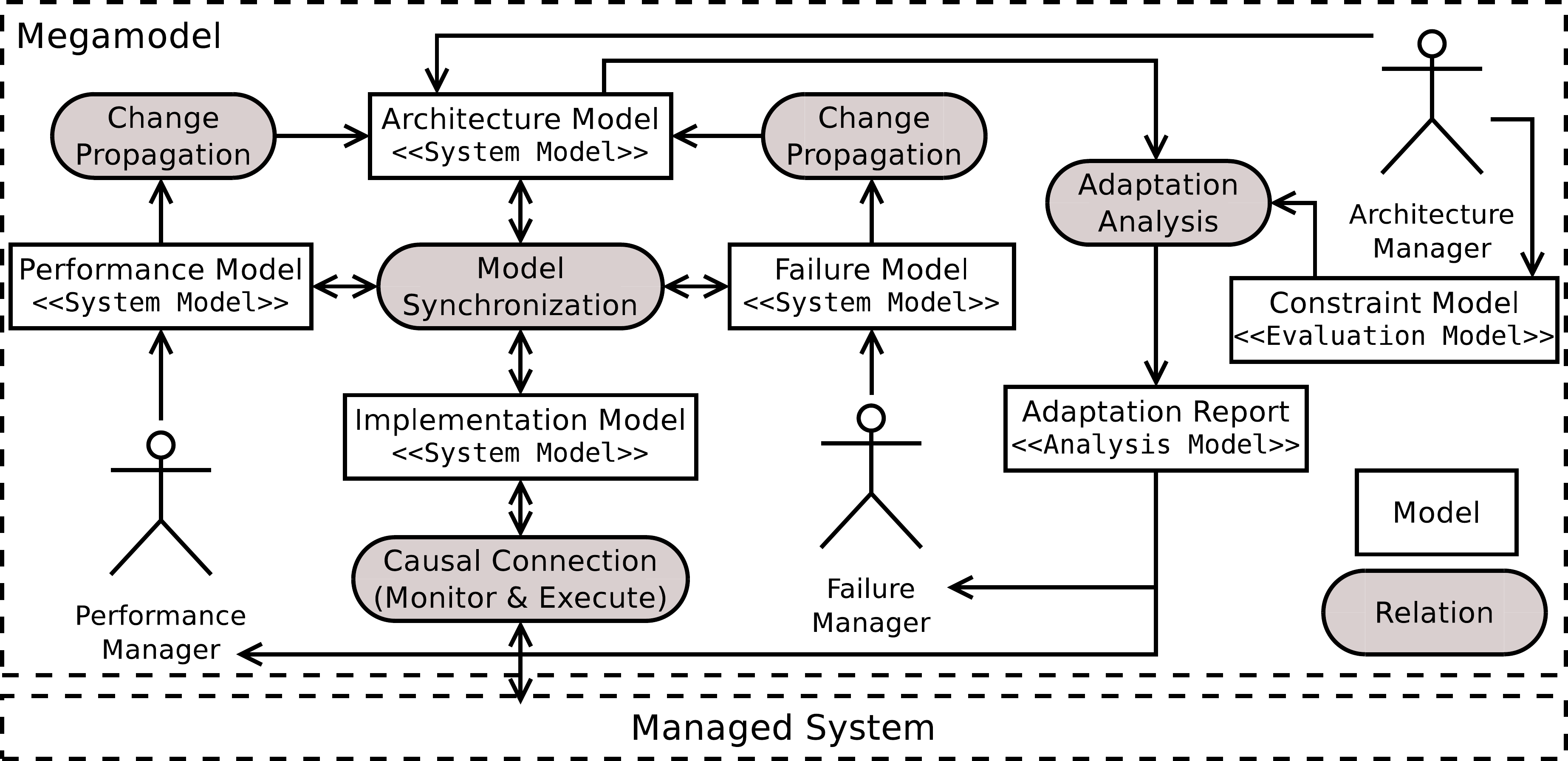}
  \caption{A Megamodel Example for a Self-Adaptive Software System}
  \label{fig:selfadapt}
\end{center}
\vspace{-9mm}
\end{figure}

In this section, we outline a case study in the field of self-adaptive software that exemplifies the role and benefits of models and megamodels at runtime. The case study is based on our previous work that employs several runtime models simultaneously for monitoring~\cite{VogelNHGB10} and adapting~\cite{VG10} a system as outlined in Figure~\ref{fig:selfadapt}. Using stereotypes, the models are mapped to the categories presented in Section~\ref{subsec:SAS+Models} while neglecting the distinction between descriptive and prescriptive models due to space constraints. The \emph{Managed System} is reflected by an \emph{Implementation Model} and both are causally connected to realize the monitoring and the execution of changes. However, the implementation model is platform-specific, complex, at a low level of abstraction, and related to the system's solution space. Therefore, abstract runtime models are derived from the implementation model using incremental and bidirectional \emph{Model Synchronization} techniques. These abstract models can be causally connected to the system via the implementation model, and they are similar to \emph{Configuration and Architectural Models} (cf. Section~\ref{subsec:models}). Each of these abstract models focuses on a specific concern of interest, which leverages models related to problem spaces. An \emph{Architecture Model}, a \emph{Performance Model}, and a \emph{Failure Model} are derived focusing on architectural constraints, performance, and failures of the system, respectively. Thus, specific self-management capabilities are supported by distinct models, like self-healing by the failure model or self-optimization by the performance model. Consequently, specialized autonomic managers, like a \emph{Performance Manager} working on the performance model, can be employed. The managers' tasks are the analysis of the system and primarily the planning of adaptations with respect to the specific concerns. 

However, adaptations planned by a certain manager due to a specific concern might interfere with other concerns covered by other managers. For example, adaptations based on the performance model, like deploying an additional component to balance the load, might violate architectural constraints covered by the \emph{Architecture Model}, like the affected component can only be deployed once.

Since each concern is covered by a different model, megamodels can be used to describe relations, like interferences or trade-offs, between the different models or concerns. This makes these relations explicit such that they can be utilized for modeling coordination mechanisms between different managers to balance multiple concerns. Besides describing these mechanisms, they can also be enacted at runtime as outlined by the following scenario. Before any adaptation planned by the performance or failure manager who change the performance or failure model, respectively, is executed on the system by triggering the \emph{Model Synchronization}, the changes are automatically propagated to the architecture model (cf. \emph{Change Propagation} relations in Figure~\ref{fig:selfadapt}). Then, the architecture manager applies the \emph{Constraint Model} on the updated architecture model to analyze the planned adaptations (\emph{Adaptation Analysis}) by writing an \emph{Adaptation Report}. This report is sent to the manager proposing the adaptation and it instructs either the execution of the planned adaptation on the system or the rollback of the corresponding model changes depending on the analysis results.

The presented case study exemplified a potential use case and benefits of megamodel concepts for self-adaptive systems in organizing and utilizing multiple runtime models and relations, especially regarding the execution of a loop.

\section{Conclusion and Future Work}\label{sec:conclusion}

In this paper we have shown that the issue of complexity in model-driven software development, caused by the amount of development models and their relations, is also a problem in the domain of runtime system management and self-adaptive systems. Since for the latter domain this problem is rather neglected by applying ad-hoc solutions, we proposed to use megamodel concepts at runtime. Therefore, we presented a categorization of runtime models and potential relations between the models, which outlined the role of models at runtime. Moreover, it demonstrated that advanced approaches already or likely use multiple models and relations simultaneously. Based on that, we showed that megamodels are an appropriate formalism to manage multiple runtime models and their relations. We especially discussed the role of interrelated models and megamodels at runtime for the case of self-adaptive systems, which was also exemplified by a case study.

The discussions at the \emph{2010 Models@run.time} workshop basically concluded that multiple runtime models are required to provide views at different levels of abstraction, for different time scales regarding the feasible performance of activities working on runtime models, and for various purposes, like monitoring, analysis, decision-making, or adaptation. These discussions further motivate our work on investigating multiple runtime models and their relations.

As future work, we plan to elaborate our categorization to incorporate other preliminary classifications comparing development and runtime models~\cite{France+Rumpe2008}, and describing dimensions of runtime models, like structural/behavioral or procedural/declarative models~\cite{MC.2009.326}. This includes possible categorizations of model relations, which requires a more profound understanding of the different kinds of runtime models and their usage. Finally, we will investigate the application of megamodel concepts in our approach~\cite{VG10,Vogel-ICAC09,VogelNHGB10}, which will potentially uncover yet unknown specifics of megamodel concepts for the case of runtime models.


\end{document}